\documentclass[epj]{svjour}

\usepackage{psfig,epsfig}
\usepackage{amsmath}	
\usepackage{amssymb}	
\usepackage{amsfonts}	

\def\dd{\ensuremath{\text{d}}}
\def\sumgamma{\ensuremath{\sum\limits}_{\gamma=\pm}}
\def\Min{\ensuremath{\mathop{\,\rm Min\,}\limits}}  
\def\bm{\beta_{-}}
\def\bp{\beta_{+}}
\def\AB{\mathrm{AB}}
\def\AC{\mathrm{AC}}
\def\BD{\mathrm{BD}}
\def\CD{\mathrm{CD}}
\def\AD{\mathrm{AD}}
\def\BC{\mathrm{BC}}
\def\A{\mathrm{A}}
\def\B{\mathrm{B}}
\def\C{\mathrm{C}}
\def\D{\mathrm{D}}
\def\M{\mathrm{M}}
\def\Ll{\mathrm{L}}
\def\LL{\mathrm{L'}}
\def\LM{\mathrm{LM}}
\def\LLB{\mathrm{L'B}}
\def\MB{\mathrm{MB}}
\def\inter{\mathrm{int}}
\def\at{\mathrm{at}}
\def\av{\mathbf{av}}
\def\ph{\mathrm{phot}}
\def\tran{\mathrm{trans}}
\def\mP{m_{\mathrm{P}}}
\def\lP{\ell_{\mathrm{P}}}
\def\tP{t_{\mathrm{P}}}
\def\lC{\ell_{\mathrm{C}}}
\def\gw{\mathrm{gw}}
\def\kB{k_{\mathrm{B}}}
\def\beps{\mathbf{e}}
\def\bx{\mathbf{x}}
\def\bk{\mathbf{k}}
\def\bK{\mathbf{K}}
\def\bn{\mathbf{n}}
\def\bu{\mathbf{u}}
\def\cA{\mathcal{A}}

\def\makeheadbox{}

\begin{document}

\title{Gravitational decoherence of atomic interferometers}
\author{Brahim Lamine \inst{1} \and Marc-Thierry Jaekel \inst{2} 
\and Serge Reynaud \inst{1}}
\institute{Laboratoire Kastler Brossel 
\thanks{Unit\'e mixte du CNRS, de l'\'Ecole Normale Sup\'erieure 
et de l'Universit\'e Pierre et Marie Curie.} , 
Universit\'e Pierre et Marie Curie, case 74,\\
Campus Jussieu, place Jussieu, F-75252 Paris Cedex 05
\and Laboratoire de Physique Th\'eorique
\thanks{Unit\'e propre du CNRS, associée \`a l'\'Ecole Normale 
Sup\'erieure et \`a l'Universit\'e Paris XI Orsay.} , 
\'Ecole Normale Sup\'erieure,\\
24 rue Lhomond, F-75231 Paris Cedex 05}
\date{Received: February 14, 2002}
%
\abstract{
We study the decoherence of atomic interferometers due to
the scattering of stochastic gravitational waves.
We evaluate the `direct' gravitational effect registered 
by the phase of the matter waves as well as the `indirect'
effect registered by the light waves used as beam-splitters
and mirrors for the matter waves.
Considering as an example the space project HYPER,
we show that both effects are negligible for the presently 
studied interferometers.
} 
\maketitle

\section{Introduction}

The idea that spacetime fluctuations could play a universal role in
the transition from quantum to classical physics has been proposed by
a number of authors. Already present in the Feynman lectures on
gravitation \cite{Feynman99,Feynman63}, it was more thoroughly
developed and popularized for instance in
\cite{Karolyhazy66,Diosi89,Penrose96,Ellis98}. An important argument
in favor of such an idea is that the Planck mass, i.e. the mass scale
$\mP =\sqrt{\frac{\hbar c}{G}} $ built on the Planck constant $\hbar$,
the velocity of light $c$ and the Newton constant $G$, has a value
$\mP \simeq 22\ \mu$g lying on the borderland between microscopic and
macroscopic masses.  In other words, microscopic masses could be
characterized as masses $m<\mP$ for which the associated Compton
length $\lC=\frac{\hbar }{mc}$ is larger than the Planck length
$\lP=\sqrt{\frac{\hbar G}{c^3}}\sim 10^{-35}$m whereas macroscopic
masses $m>\mP$ would correspond to a Compton length $\lC$ smaller than
the Planck length $\lP$.

Clearly, this dimensional argument is not by itself sufficient to
reach definitive conclusions. It can be hoped that the existence of
fundamental spacetime fluctuations, with a length scale determined by
$\lP$, may be revealed by long-term diffusion effects in the same
manner as microscopic molecular motion is revealed by Brownian motion.
It has for instance been proposed that intrinsic spacetime
fluctuations could be observed through a decoherence effect which
might be visible with matter-wave interferometers
\cite{Percival97,PercivalS97,Power00,Amelino99,Amelino00}.  The effect
has not been seen in existing matter-wave interferometers
\cite{Peters99,Peters01,Gustavson00}. More sensitive instruments are
being developed, like the atomic interferometer HYPER designed to
measure the Lense-Thirring effect in a space-borne experiment, and it
is important to estimate the ultimate decoherence due to fundamental
spacetime fluctuations for such an instrument \cite{Hyper00}.

The aim of the present paper is to give quantitative answers to this
question by considering the decoherence mechanism associated with the
scattering of gravitational waves present in our celestial
environment.  These gravitational waves are the intrinsic fluctuations
of spacetime predicted by general relativity \cite{Misner}.  The
latter theory can be used as an accurate effective theory of
gravitation for all frequencies ever explored in experiments
\cite{Will90,Damour94}. The gravitational waves are the intrinsic
field fluctuations predicted by the linearized form of the theory
\cite{Weinberg65,Grishchuk77,Zeldovich86}. This linearized form is
widely used for studying propagation of gravitational waves and their
interaction with the presently developed interferometric detectors
\cite{Schutz99,Maggiore00,Ungarelli00}.

In the present paper, we will study the decoherence of an atomic
interferometer due to its interaction with the stochastic background
of gravitational waves emitted by unresolved sources in our galaxy or
its vicinity \cite{Hils90,Giazotto97}. These waves dephase differently
the matter-waves on the two interfering paths as well as the
light-waves used to build up beam-splitters and mirrors in atomic
interferometers \cite{AtomInt97}. When averaged over the integration
time of the measurement, the differential dephasing results in a loss
of contrast of the interference fringes.  We will use the remaining
fringe contrast to characterize the decoherence and write it in terms
of the geometry of the interferometer and of the statistical function
describing the gravitational environnement.

In this approach, decoherence will be understood as resulting from a
phase dispersion due to the unobserved degrees of freedom of the
gravitational environnement. To be more precise, gravitational waves
with frequencies in the detection window of the interferometer have to
be interpreted as signals while frequencies outside the detection
window are ignored.  In a typical situation, the detection window
corresponds to frequencies smaller than the inverse of an averaging
time.  The integration over frequencies outside the detection window
can thus be identified with the trace over the environnemental degrees
of freedom usually considered in theoretical studies of decoherence
\cite{Zeh70,Dekker77,Zurek81,Zurek82,Caldeira83,Caldeira85,Joos85,Raimond01}.
The phase dispersion approach used in the present paper is known to be
equivalent to the other approaches to decoherence \cite{Imry90} and it
is obviously well adapted to the description of interferometers where
the phase is the natural variable.

We will show in this paper that the scattering of gravitational
background does not lead to an appreciable decoherence effect for the
atomic interferometers presently studied, HYPER being chosen as the
typical example. Incidentally, this means that atomic interferometers
will not have their interference fringes destroyed by this decoherence
mechanism. This answer has to be contrasted with recently published
results which prove that the scattering of stochastic gravitational
waves present in our galactic environment is the dominant, and
extremely efficient, decoherence mechanism for macroscopic motions,
say the planetary motion of the Moon around the Earth
\cite{Reynaud01,Reynaud02}.  This contrast is easily explained by the
already evoked dimensional argument~: gravitational decoherence
effects are likely to be more efficient for macroscopic masses than
for microscopic objects. In particular, the mass of the Moon is larger
than Planck mass by orders of magnitude whereas the microscopic
entities used as spacetime probes in atomic interferometers have their
mass much smaller than Planck mass. However, as already stated, this
simple scaling argument is not by itself sufficient to answer
quantitative questions about the decoherence rates. In the following,
we will give precise estimations of the decoherence effect which
depend not only on the mass of the atoms, but also on their velocity,
on the geometry of the interferometer and on the noise spectrum
characterizing the gravitational background in the relevant frequency
range.

\section{Gravitational backgrounds}

A first step is to characterize the fundamental fluctuations of
spacetime and their effect on the motion of matter. Although a
complete quantum theory of gravity is not available, it is possible to
describe spacetime fluctuations in our environment. At the frequencies
of experimental interest, which are much smaller than Planck
frequency, they are identified \cite{Reynaud01,Reynaud02} as the
stochastic backgrounds of gravitational waves currently studied in
relation with the development of gravitational wave detectors
\cite{Schutz99}.

The effect of gravitational perturbations may in principle be
described in a manifestly gauge-invariant manner. In the present
paper, we will adopt the common strategy of studies of gravitational
waves~: admitting that this point can be dealt with, we then chose a
specific gauge, namely the transverse traceless (TT) gauge with metric
perturbations differing from zero only for purely spatial components
$h_{ij}$ ($i,j=1,2,3$ stand for the spatial indices whereas $0$
represents the temporal index).  Then gravitational waves are
conveniently described through a mode decomposition
\cite{Blanchet00}~:
\begin{eqnarray}
&&h_{ij}(x)=\int \frac{\dd^4 k}{(2\pi)^4}h_{ij}[k] e^{-ik_{\mu}x^{\mu}} 
\nonumber \\
&&h_{00} = h_{0i}=0
\label{decompo1}
\end{eqnarray}
Any Fourier component is a sum over the two circular polarizations~:
\begin{equation}
h_{ij}[k]= \sumgamma \left( 
\frac{\beps ^{\gamma}_i[k] \beps ^{\gamma}_j[k]}{\sqrt{2}} 
\right) ^* h^{\gamma}[k]
\label{decomposition}
\end{equation}
Gravitational waves correspond to wavevectors $k$ lying on the light
cone and they are transverse with respect to this wavevector~:
\begin{eqnarray}
&&k^2 = k_0^2-\bk^2= 0 \quad , \quad k_0 \equiv \frac{\omega}{c} 
\nonumber \\
&&\bk_i h_{ij}=0
\end{eqnarray}
The gravitational polarization tensors are obtained as products of the
polarization vectors $\beps^\pm$ well-known from electromagnetic
theory.  When necessary, we will chose the following representation
for the unit vector $\bn $ along the propagation direction of the
gravitational wave and the corresponding polarization vectors~:
\begin{eqnarray}
&&\bn   \equiv \frac{c\bk}\omega 
= \left( \begin{array}{l}
\sin\theta\cos\varphi \\
\sin\theta\sin\varphi\\
\cos\theta
\end{array}\right) \quad , \quad \omega>0 \nonumber \\
&&\beps ^{\gamma} [k] =\left( \begin{array}{c}
-\cos\theta\cos\varphi+i\gamma\sin\varphi \\
-\cos\theta\sin\varphi-i\gamma\cos\varphi \\
\sin\theta \end{array}\right)
\label{polariz}
\end{eqnarray}
Spatial vectors are written as bold letters. Note that reality
conditions for the perturbation metric $h_{ij}(x)$ are read~:
\begin{eqnarray}
&&\big(\beps _i^{\gamma}[-k]\big)^*=\beps _i^{\gamma}[k]
\nonumber \\
&&\big(h^{\gamma}[k]\big)^*=h^{\gamma}[-k]
\end{eqnarray}

We use the natural caracterization of the stochastic background in
terms of the spectral density of strain fluctuations $C_{hh}[k]$ of
the metric. For simplicity, we consider the case of gaussian,
stationary, unpolarized and isotropic backgrounds~:
\begin{equation}
\left\langle h^{\gamma}[k]h^{\gamma'}[k']\right\rangle=(2\pi)^{4}\,
\delta^{\gamma\gamma'}\,\delta^4 (k+k')\, C_{hh}\left[ k\right] 
\label{correlation}
\end{equation}
The general case could be dealt with by considering arbitrary
correlations between amplitudes $h^{+}\left[ k\right] $ and
$h^{-}\left[ k\right] $. This would allow one to take into account
polarized or anisotropic backgrounds, as is necessary for a thorough
analysis of the galactic background, as well as non stationnary fields
appearing in some cosmological models \cite{Grishchuk77,Grishchuk90}.

Gravitational backgrounds are usually written in terms of one metric
component (say $h_{12}$, but the result would be the same for other
components due to the isotropy assumption) at a fixed spatial position
(say $\bx =0$, but the result would be the same for other positions
due to the stationarity assumption) as a function of time $t$. They
are thus described by the spectral density $S_h [\omega]$ of strain
fluctuations considered in most papers on gravitational waves
detectors \cite{Maggiore00}~:
\begin{equation}
\left< h_{12}(t)h_{12}(0)\right>=\int\frac{\dd\omega}{2\pi}S_h [\omega]
e^{-i\omega t}
\label{chh}
\end{equation}
This noise spectrum, written in the TT gauge, is not gauge invariant.
This is not a problem since only gauge invariant quantities will be
computed in the following. It follows from equations
(\ref{chh},\ref{decomposition}) that $S_h [\omega]$ is obtained by
integrating $C_{hh}[k]$ over the momenta $k$ which correspond to a
given frequency $\omega$~:
\begin{align}
&S_h [\omega]=\int \frac{\dd |\bk |}{4\pi^2c} \sumgamma 
\big<\beps _1^{\gamma}[k]\beps _2^{\gamma}[k]\beps _1^{\gamma}[-k]
\beps _2^{\gamma}[-k]\,C_{hh}[k]
\big>_{\bn }
\nonumber \\
&\left<f[\bn ]\right>_{\bn }=\int\frac{ \dd^2\bn }{4\pi}\;f[\bn ]
\quad , \quad \dd^2\bn \equiv \dd\cos\theta \dd\varphi
\end{align}
We have denoted by $<\dots>_\bn$ the averaging over spatial directions
of wavevectors at a given frequency.  Using the fact that
gravitational wavevectors lie on the light cone and the simplifying
assumptions already described, we inverse the preceding relation to
obtain~:
\begin{equation}
C_{hh}[k]= 10\pi^2c^2 \delta(k^2) \frac{S_h [\omega]}{\omega}
\label{sh} 
\end{equation}

In this paper, we will consider the binary confusion background
describing gravitational waves emitted by unresolved binary systems in
our galaxy or its vicinity. This background is represented for example
on figure (1) of reference \cite{Schutz99}. It relies on the laws of
physics and astrophysics as they are known in our local celestial
environment. In particular, it depends on the statistical repartition
of binary systems in the sky. In the frequency range of interest,
which is discussed in more detail below, the binary confusion
background dominates other sources of stochastic gravitational waves,
in particular those associated with cosmological contributions. Note
that this frequency range, between $\mu$Hz and mHz, corresponds to
frequencies much smaller than the detection window of VIRGO and other
ground-based optical interferometers but roughly of the same order as
the frequencies in the detection window of the space-borne project
LISA
\footnote{Informations on the ground-based GW detectors may be found
on the Web sites; VIRGO~: http://www.virgo.infn.it/~; GEO~:
http://www.geo600.uni-hannover.de/~; LIGO~:
http://www.ligo.caltech.edu/~; TAMA~:
http://tamago.mtk.nao.ac.jp/tama.html~; ACIGA~:
http://www.anu.edu.au/Physics/ACIGA/~.
\newline
Informations on the space-borne interferometer LISA may be found on
the Web sites at NASA~: http://lisa.jpl.nasa.gov/ and ESA~:
http://sci.esa.int/home/lisa/~.}.

The properties of the gravitational bath may also be characterized by
an effective number $n_\gw$ of gravitons per mode (the precise
relationships between $n_\gw[\omega]$ and $S_h [\omega]$ are given in
the last section). It is worth noting immediately that this number
$n_\gw$ is extremely large, so that the gravitational environment
corresponds to the limit of high-temperature classical fluctuations.
As a consequence, the vacuum fluctuations of the gravitational field
\cite{Grishchuk90}, which have been shown to lead to ultimate
fluctuations of geodesic distances of the order of Planck length
\cite{Jaekel94,Jaekel95QSO,Jaekel95AP}, are ignored in the present
paper. We also note that, as the number of unresolved binary systems
contributing to the binary confusion background is large and as these
sources are independent of each other, it appears quite safe to
consider that the gravitational background obeys gaussian statistics
(this might not be true for cosmological contributions).

In order to discuss our main results in the end of the present paper,
we will use the fact that the binary confusion background corresponds
to a nearly flat function $S_h [\omega]$ in the frequency range
$1-100\mu$Hz.  This means that the gravitational noise spectrum is
quasi-thermal at such frequencies and entails that the decoherence
mechanism can be interpreted as a Brownian-like diffusion process. We
will estimate quantitatively the quantity playing the role of the
diffusion coefficient and show that it does not lead to an appreciable
decoherence effect for atomic interferometers like HYPER.

\section{Gravitational decoherence}

Now we want to quantify the effect of the previously discussed
gravitational waves on the coherence properties of an atomic
interferometer.  In the present section, we present a detailed
discussion of this effect which could basically be schematized as
follows~: coherence of the interference fringes is preserved if and
only if the differential phase perturbation between the two arms is
controlled to a level much better than $2\pi$.

When propagating in spacetime, the atomic probe field registers
curvature fluctuations. The main effect of the perturbation is
described by the eikonal approximation, valid for wavevectors $k$ of
the gravitational wave much smaller than wavevectors $K$ of the probe
field.  This effect is characterized as a dephasing $\Phi$ of the
probe field, evaluated at a fixed spatial position (say $\bx =0$) as a
function of time $t$. At the lowest order, the dephasing is linear in
the metric and can be decomposed over the gravitational wave modes~:
\begin{equation}
\Phi(t)=\int\frac{\dd^4 k}{(2\pi)^4} 
\sumgamma \phi_k^{\gamma}(t) h^{\gamma}[k]
\end{equation}
At the moment, $\phi_k^{\gamma}(t)$ are time-dependent coefficients
which depend on the geometry of the interferometer. These coefficients
will be explicitly written in the forthcoming sections but we already
know that they satisfy the reality conditions~:
\begin{equation}
\phi_{-k}^{\gamma}=\left( \phi_k^{\gamma} \right)^* 
\label{realitycond}
\end{equation}

For a stationary, isotropic and unpolarized background, the
correlation function for the dephasing is deduced~:
\begin{equation}
\left< \Phi(t)\Phi(t')\right>=\int\frac{\dd^4k}
{(2\pi)^4}\sumgamma \phi_k^{\gamma}(t) \phi_{-k}^{\gamma}(t')
C_{hh}[k]e^{-i\omega (t-t')}
\end{equation}
The gravitationally induced dephasing is not always a stationary
noise.  Here, we will focus our attention on situations where it is
stationary or quasi-stationary and where the correlation function is
simply represented by a noise spectrum~:
\begin{eqnarray}
\left< \Phi(t)\Phi(t')\right>&=&\int \frac{\dd \omega}{2\pi}
S_\Phi [\omega] e^{-i\omega (t-t')}
\nonumber \\
S_\Phi [\omega]=S_h [\omega] \cA [\omega] \;&,&\;
\cA [\omega]=\frac{5}{2}
\sumgamma \left< |\phi_k^{\gamma}|^2\right>_{\bn }
\end{eqnarray}
In fact, we have supposed that the amplitudes $\phi_k^{\gamma}(t)$ are
only slowly dependent functions of time and, furthermore, we have used
the reality condition (\ref{realitycond}).  The equation thus obtained
means that the stochastic dephasing has a noise spectrum equal to the
product of two factors, the gravitational noise spectrum $S_h
[\omega]$ and the apparatus response function $\cA [\omega]$ which
depends on the geometry of the apparatus.

As mentioned earlier, the analysis of decoherence should in principle
take into account the detection strategy used in the interferometric
measurement. In order to fix ideas, we will consider a simple strategy
where the output of the interferometer, supposed to be linear in the
variation of the dephasing $\Phi$, is averaged over an averaging time
$\tau_\av$. This means that the signal, for instance the
Lense-Thirring effect in HYPER experiment \cite{Hyper00}, is contained
in the averaged dephasing $\overline{\Phi}$ which is read in the
frequency domain as~:
\begin{equation}
\overline{\Phi} [\omega] = \frac{\Gamma}{\Gamma - i\omega} \Phi [\omega] 
\quad , \quad \Gamma \equiv \frac {1}{\tau_\av}
\end{equation}
In other words, the signal window is defined by a low-pass filter with
a bandwidth $\Gamma$. Now, the frequencies outside the signal window
constitute an uncontrolled noise which may degrade the coherence of
the interferometer if it is large enough.

In order to estimate this potential decoherence effect, we define the
uncontrolled dephasing which is also the dephasing after a high-pass
filter with the same bandwidth~:
\begin{equation}
\delta \Phi [\omega] = \Phi [\omega] - \overline{\Phi} [\omega] = 
\frac{-i \omega}{\Gamma - i\omega} \Phi [\omega] 
\end{equation}
We then consider the visibility $\mathcal{V}$ of the fringes which is
the mean value of the exponential of the uncontrolled phase noise~:
\begin{eqnarray}
\mathcal{V} &=& \big<\exp \left( i \delta \Phi(t) \right) \big> 
\end{eqnarray}
As we have supposed the gravitational background to obey the gaussian
statistics, the various dephasings are also gaussian stochastic
variables.  It follows that the visibility $\mathcal{V}$ of the
fringes may be expressed in terms of the variance of the uncontrolled
noise~:
\begin{eqnarray}
&&\mathcal{V} = \exp \left( -\frac{\Delta\Phi^2}{2} \right) 
\quad , \quad \Delta \Phi^2 = \left< \delta \Phi(t) \delta \Phi(t) 
\right> 
\label{contrast}
\end{eqnarray}
Finally this variance is given by the following integral over
frequency~:
\begin{eqnarray}
\Delta \Phi^2 &=& \int\frac{\dd\omega}{2\pi} \, S_h [\omega] \, 
\cA [\omega] \, \frac{\omega^2}{\omega^2+\Gamma^2}  
\label{eqfond}
\end{eqnarray}

The last two equations are a fondamental result of this article. They
quantitatively characterize the gravitational decoherence effect
through the reduction of the fringe visibility $\mathcal{V}$.  This
visibility is the exponential of the phase noise variance.  The latter
is an integral over the whole frequency spectrum of a product of
factors. Besides the gravitational noise spectrum $S_h [\omega]$ and
the apparatus response function $\cA [\omega]$ which have already been
discussed, there is a further factor, a high-pass filter with a cutoff
$\Gamma$, which defines uncontrolled noise as corresponding to
frequencies outside the detection window. For simplicity, we have
considered here a Lorentzian expression associated with a simple
averaging strategy.  Note that the filter may help us to regularize
potential infrared divergences in the forthcoming calculations.  We
shall see later on that this is not necessary for the situations
considered in this paper and that the results are essentially
independent of the cutoff. Note also that it would be easy to replace
the Lorentzian filter by more sophisticated expressions corresponding
to different signal detection strategies.

As it is usual in decoherence theory, the remaining coherence, here
the visibility $\mathcal{V}$, is the exponential of a noise variance,
the variance $\Delta\Phi^2$ of the uncontrolled noise. From the point
of view of the interferometrist, this expression has a quite clear
expression~: the gravitational perturbation dephases differently the
waves in the two arms, and the fringe contrast is appreciably degraded
if and only if the resulting variance $\Delta\Phi^2$ is of the order
or greater than unity.  At this point, it is worth emphasizing that
the same result would have been obtained through more formal
approaches of decoherence with the uncontrolled noise frequencies thus
interpreted as the degrees of freedom of the environment (see for
example \cite{Imry90} for a more detailed discussion of this point).

Expression (\ref{eqfond}) is valid for a gaussian, stationary,
isotropic and unpolarized background. The numerical factor $\frac 52$
is a factor arising from angular averaging for an isotropic and
unpolarized background. As already mentioned, generalization to
anisotropic, polarized and non stationary background is possible. Even
non gaussian noise can in principle be dealt with by developping the
fringe visibility in cumulants of the stochastic noise. However, these
refinements would not change the main conclusion of the paper, namely
that the gravitational decoherence is not efficient for presently
studied atomic interferometers such as HYPER.  The next sections are
devoted to explicit calculations of the apparatus function $\cA
[\omega]$ for such an interferometer.

\section{Atomic interferometer with a Mach-Zehnder geometry}

In order to smoothly introduce the more technical parts of our
evaluations, we first consider in this section an hypothetical atomic
interferometer with the same Mach-Zehnder geometry as HYPER but where
optical elements, beam splitters and mirrors, would be built up from
massive and motionless material pieces.  A more realistic description
of HYPER with optical elements built up on stimulated Raman processes
is presented in the next section.

\begin{figure}[htbp]
$$
\epsfig{file=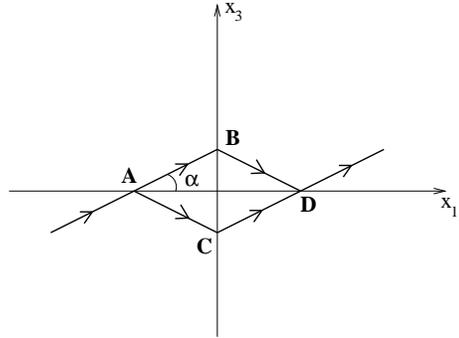,width=6cm}
$$
\caption{Schematic description of an atomic interferometer with a
Mach-Zehnder geometry~: thanks to the presence of beam splitters at A
and D and mirrors at B and C, atomic matter-waves are coherently
recombined after having propagated along the two geodesic paths ABD
and ACD; the output intensity is then a sinusoidal function of the
differential dephasing $\Phi$ between the two paths. For simplicity,
we consider that the interferometer has the symmetry of the rhomb.
The angle $2\alpha$ is exaggerated on the figure.}
\label{zehnder}
\end{figure}

The Mach-Zehnder geometry is represented on figure (\ref{zehnder}).
The atomic matter-waves are supposed to follow their classical
trajectories. Equivalently, they obey the propagation equation of a
scalar field with a wavevector $K_{\mu}$ and the dephasing $\Phi_\at $
is the first-order effect of the gravitational perturbation of the
associated lagrangian. Then, the dephasing evaluated on a closed loop
is gauge-invariant, as soon as conservation of the energy-momentum
tensor is properly taken into account \cite{Landau}. This raises
specially delicate problems for the description of beam splitters and
mirrors in the interferometer.

In the present section, we solve these problems by using the method
commonly adopted for the description of optical interferometers
\cite{Grishchuk88,Hellings92}~: the treatment is largely simplified
when optical elements such as beam splitters and mirrors are described
as heavy objects initially at rest and gravitational waves described
in the TT gauge. In this gauge indeed, gravitational waves have no
effect on massive objects at rest. And objects with a large mass may
be considered as staying at rest when initially at rest, since the
momentum transfered by the field upon scattering does correspond to a
negligible velocity change. This implies that the effect of
gravitational waves on optical elements can be ignored throughout the
calculation.  This treatment is certainly an approximated one. In
particular, the assumptions just discussed entail that the
interferometer is at rest. This is obviously not the case for HYPER
which is orbiting around the Earth, this motion being important for
the analysis of the signal associated with the looked for
Lense-Thirring effect \cite{Hyper00}.  However this motion may be
disregarded in the discussion of decoherence in the present paper. In
the present section, we build up our calculations on this
approximation. 

Now we write the dephasing $\Phi_\at (t)$ obtained from the geodesic
deviation equation for the probe field~:
\begin{equation}
\Phi_\at (t)=\frac{K_0}{2}\int_{t}^{t+\tau} \! h_{ij}(t')
\bu ^i(t') \bu ^j(t')c\dd t' \label{dephasing}
\end{equation}
This equation is written in the eikonal approximation where the
wavevector $k_{\mu}$ of the gravitational waves is much smaller than
the wavevector $K_{\mu}$ of the probe field \cite{Mashhoon80}.  The
integral (\ref{dephasing}) is taken along the unperturbed geodesic
path $x(t')$ of the probe with the coordinate time $t'$ used as an
affine parameter; $\tau$ denotes the time of propagation along the
path, and $\bu $ is the reduced wave vector~:
\begin{equation}
\bu  = \frac{\bK}{K_0}
\end{equation}
We then deduce the coefficients $\phi_k^{\gamma}(t)$ which
characterize the sensitivity of the dephasing to the gravitational
wave $h^{\gamma}[k]$~:
\begin{equation}
\phi_k^{\gamma}(t)=\frac{K_0}{2\sqrt{2}}\int_{t}^{t+\tau}\!\!
\left( \beps ^\gamma [k] ^* \mathbf{.} \bu (t') \right) ^2
e^{-ik_\mu x^\mu (t')}c\dd t'
\label{phikgamma}
\end{equation}
These coefficients are time-independant, due to the stationarity
assumption and to the fact that we have supposed the interferometer to
be at rest.

As a consequence of the preceding discussions, the global dephasing of
the interferometer is obtained by adding the contributions of the free
propagation segments~:
\begin{eqnarray}
\phi_k^{\gamma}&=&\phi_k^{\gamma}[\AB]+\phi_k^{\gamma}[\BD]
-\phi_k^{\gamma}[\AC]-\phi_k^{\gamma}[\CD] 
\end{eqnarray}
We ignore any dephasing of the probe field associated with scattering
on the beam splitters. This point will be discussed in more details in
the next section.

For simplicity, we consider the interferometer to have a rhombic
geometry with $2\alpha$ the aperture angle and $\ell_\AB$ the length
of each side.  The latter quantity is related to the time of flight
$\tau_\AB$ and the modulus $v_\at$ of the atomic velocity~:
\begin{equation}
\ell_\AB = v_\at \tau_\AB   
\end{equation}
We suppose the interferometer to lie in the plane $(\bx ^1, \bx ^3)$
and write the coordinates of its apexes in a manner exploiting the
symmetry of the rhomb ($\bx ^2=0$ for all points)~:
\begin{eqnarray}
t_\D=\tau_\AB \quad &,& \quad 
\bx_\D=\big(\ell_\AB\cos\alpha,\,0,\,0 \big) \nonumber \\
t_\A=-t_\D \quad &,& \quad \bx_\A=-\bx_\D \nonumber \\
t_\B=0 \quad &,& \quad 
\bx_\B = \big(0,\, 0,\, \ell_\AB\sin\alpha \big) \nonumber \\
t_\C=t_\B \quad &,& \quad \bx_\C=-\bx_\B
\end{eqnarray}
We have used the stationarity to select a specific time.  Then the
reduced velocities are read~:
\begin{eqnarray}
&&\bu_\AB = \frac{v_\at}{c} \big( \cos\alpha ,\, 0 ,\, \sin\alpha \big)
\nonumber \\
&&\bu_\BD = \frac{v_\at}{c} \big( \cos\alpha ,\, 0 ,\, -\sin\alpha \big)
\nonumber \\
&&\bu _\BD = \bu _\AC 
\quad , \quad \bu _\CD = \bu _\AB 
\end{eqnarray}

We now evaluate the amplitude $\phi_k^{\gamma}[\AB]$ corresponding to
the segment [AB] from equation (\ref{phikgamma}). We restrict our
attention to the non relativistic limit $v_\at \ll c$ so that~:
\begin{equation}
\phi_k^{\gamma}[\AB]\simeq\frac{i}{2\sqrt{2}} \frac{mc^2}{\hbar\omega}
\left( \beps^{-\gamma} \mathbf{.} \bu_\AB \right)^2 \left( e^{-i\omega t_\B} 
- e^{-i\omega t_\A} \right)
\end{equation}
Using the symmetry of the rhomb, we then deduce the amplitude
$\phi_k^{\gamma}$ corresponding to the sum over the four segments~:
\begin{eqnarray}
\phi_k^{\gamma}&=& \frac{2i}{\sqrt{2}}\frac{mc^2}{\hbar\omega}
\left( \left( \beps^{-\gamma} \mathbf{.}  \bu _\AB \right)^2 
- \left( \beps^{-\gamma} \mathbf{.} \bu _\AC \right)^2 \right)
\sin^2\frac{\omega\tau_\AB}{2}  \nonumber \\
&=& 2i \sqrt{2} \frac{\Omega_\at \sin(2\alpha)}{\omega}
\beps_1^{-\gamma} \beps_3^{-\gamma} 
\left( 1 - \cos(\omega\tau_\AB) \right) \nonumber \\
\Omega_\at &=&\frac{m v_\at ^2}{2\hbar}
\end{eqnarray}
where $\Omega_\at $ is the kinetic energy of the atom measured as a
frequency. We finally get the apparatus response function~:
\begin{eqnarray}
\cA_\at [\omega] &=& 4 \frac{\Omega_\at^2 \sin^2(2\alpha)}{\omega^2}
f^2 (\omega\tau_\AB)
\end{eqnarray}
where we have introduced the function~: 
\begin{eqnarray}
f(x) &=& 2 \left( 1 - \cos(x) \right) = 4 \sin^2\frac{x}{2}  
\label{definif} 
\end{eqnarray}

We now come back to the phase noise variance (\ref{eqfond}) which
determines the fringe contrast (\ref{contrast}).  This expression
characterizes the decoherence of the atomic interferometer for an
arbitrary gravitational noise spectrum $S_h [\omega]$, for example the
spectrum describing the binary confusion background
\cite{Schutz99,Maggiore00}.  As already mentioned, we see that this
filter may help us to regularize potential infrared divergence of this
spectrum, besides the factor $f^2(\omega\tau_\AB)$ which already cuts
off low frequencies.

This discussion can be made more explicit by considering the specific
case where the spectrum is constant in the domain relevant for
evaluating the integral (\ref{eqfond}).  Note that this is
approximately the case for evaluating the effect of the binary
confusion background for HYPER. As a matter of fact, the signal has to
be integrated over an averaging time ranging between 1 day and 1
month.  Such an averaging time corresponds to the frequency range
1-30$\mu$Hz.  In this frequency range, the binary confusion background
dominates the other contributions, in particular cosmological ones,
and it has a quasi-thermal spectrum.  Hence, the integral
(\ref{eqfond}) can be estimated, at least roughly, by replacing the
frequency dependent noise spectrum $S_h [\omega]$ by a constant.
Since the binary confusion background decreases at higher frequencies,
the result obtained in this manner has to be considered as an upper
limit for the phase noise variance.

In this simple case, the integral (\ref{eqfond}) may be deduced from
the following properties of the function $f$~:
\begin{eqnarray} 
&& f^2 (x) = 4 f(x) - f(x) \nonumber \\
&& \int\frac{\dd\omega}{2\pi}\frac{f(\omega \tau)}{\omega^2+\Gamma^2}=
\frac{1-e^{-\Gamma |\tau|}}{\Gamma} 
\label{propf} 
\end{eqnarray}
so that~:
\begin{equation}
\frac{\Delta\Phi_\at ^2}{2}= 2 S_h \Omega_\at ^2 \sin^2(2\alpha) 
\frac{3-4e^{-\Gamma\tau_\AB}+e^{-2\Gamma\tau_\AB}}{\pi \Gamma} 
\end{equation}
Now, the time of flight $\tau_\AB$ of the atoms on the segment [AB] is
of the order of 1.5 s in HYPER \cite{Hyper00} and the averaging time
$\tau_\av$ is much longer. This entails that the preceding formula may
be simplified by taking the further limit~:
\begin{eqnarray}
\tau_\AB \ll \tau_\av &&\qquad \Gamma \tau_\AB \ll 1
\end{eqnarray}
with the result~:
\begin{eqnarray}
\frac{\Delta\Phi_\at ^2}{2} &\simeq& 
\frac{4}{\pi}\Omega_\at^2 \sin^2(2\alpha) S_h  \tau_\AB  
\label{decoh}
\end{eqnarray}
We note that $\sin(2\alpha)$ is the geometrical characteristic of the
interferometer which determines the difference between the two paths
and plays the role of the classicality parameter entering usual
expressions of decoherence rates \cite{Zurek81,Zurek82}.

The cutoff $\Gamma$ is no longer present in the estimate
(\ref{decoh}).  This argument may be laid down in a more general
manner.  High sensitivity measurements usually require long
integration times so that the condition $\Gamma\tau_\inter \ll 1$ is
met with $\tau_\inter $ the time of flight of the probe field in the
interferometer.  If the noise spectrum $S_h [\omega]$ has such a
low-frequency behaviour that no regularization of the integral is
needed, then the variance of the phase noise may be evaluated by
forgotting the filter~:
\begin{equation}
\Delta\Phi ^2 \simeq \int\frac{\dd\omega}
{2\pi}\:S_h [\omega] \cA [\omega]
\label{deltaphi}
\end{equation}
If, furthermore, the spectrum is nearly flat in the frequency domain
determined by the apparatus function $\cA [\omega]$, then the variance
of the phase noise is simply the product of the constant value of
$S_h$ by the integral of this apparatus function and the latter is
obtained from~:
\begin{eqnarray} 
&& \int\frac{\dd\omega}{2\pi}\frac{f(\omega \tau)}{\omega^2}= |\tau|
\label{propff} 
\end{eqnarray}

Now we proceed to a numerical evaluation of the decoherence effect.
We use the following numbers which correspond to HYPER with the choice
of Cs atoms \cite{Hyper00}~:
\begin{eqnarray}
&&m \simeq 133 \: \mbox{a.u.} \simeq 2 \times 10^{-25} \:\mbox{kg}
\nonumber \\
&& v_\at  \simeq 0.2 \: \mbox{m.s}^{-1}  \nonumber \\
&&\Omega_\at  \simeq 4 \times 10^{7} \: \mbox{Hz}  \nonumber \\
&&\sin (2\alpha) = \frac{ v_\tran }{ v_\at } \simeq 0.035  \nonumber \\
&&\tau_\AB  =\frac{\ell_\AB  }{ v_\at } \simeq 1.5 \: \mbox{s}
\end{eqnarray}
$ v_\tran $ is the transverse velocity communicated to the atoms by
the beam splitters and mirrors; it will be discussed in more details
in the next section.  Using the noise level in the frequency range of
interest $S_h \simeq 10^{-34} \mbox{Hz}^{-1}$, we finally obtain the
phase noise variance due to the dephasing of the atomic matter-waves
in HYPER~:
\begin{equation}
\frac{\Delta\Phi_\at ^2}{2} \simeq 10^{-21}
\label{odg}
\end{equation}
Clearly, the decoherence computed in this manner is completely
negligible, and the fringe contrast is unaffected by the direct
coupling of gravitational waves to atomic matter-waves.

\section{HYPER-like interferometers}

We have already noticed that, in the most sensitive presently studied
atomic interferometers, optical elements are built up on stimulated
Raman processes \cite{Peters01}. As a consequence, the dephasing seen
by the interferometer also picks up the gravitation perturbation of
the lasers involved in these optical elements. In the present section,
we give a precise evaluation of this `indirect' effect and show that
it largely dominates the `direct' atomic effect studied in the
preceding section. In order to fix the orders of magnitude, we will
still consider the numbers corresponding to the project HYPER
\cite{Hyper00}.

The use of stimulated Raman processes to build up optical elements
such as beam splitters or mirrors for atomic matter waves has been
described in a number of papers (see for example
\cite{Borde89,Borde92,Kasevich92,Weiss94,Storey94}).  In the
stimulated Raman process, atoms interact with two counter-propagating
lasers with slightly detuned frequencies $\Omega_1$ and
$\Omega_2$. The detuning is chosen so that the Raman process,
absorption of one photon in one beam and stimulated emission of one
photon in the other beam, is resonant with a transition between two
hyperfine ground states. In contrast, the detuning between one-photon
interaction and the intermediate excited state is sufficiently large
so that spontaneous emission plays a negligible role. This also
entails that the equivalent duration of the whole Raman process is so
short that the Raman process can be considered as spatially and
temporally localized.  Besides the transition from one ground level to
the other, the main effect of the Raman process is a momentum transfer
between the field and atom. This momentum transfer has its direction
along the transverse direction of the rhomb and it is responsible for
the beam splitting effect with a change of transverse velocity of the
atoms~:
\begin{equation}
v_\tran  = \frac{2 \hbar\Omega_\ph }{mc}
\end{equation}
where $\Omega_\ph$ is the nearly common value of the laser
frequencies.

\begin{figure}[htb]
$$
\epsfig{file=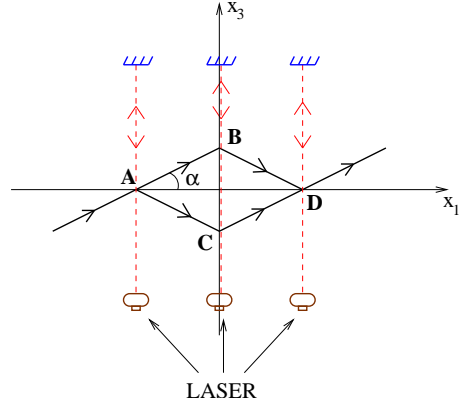,width=6cm}
$$
\caption{Schematic description of an atomic interferometer with
a Mach-Zehnder geometry and beam splitters and mirrors for atomic 
waves built up on stimulated Raman processes~: the momentum
transfer between atoms and photons produces the beam splitting;
the gravitational perturbation $\Phi$ of the dephasing between 
the two arms is now picked up not only by matter-waves but also 
by photons.}
\label{hyper}
\end{figure}

The whole interferometer is sketched on figure (\ref{hyper}) with the
atomic and photonic paths now represented. Each momentum transfer is
accompanied by a change of atomic ground state and the atoms are in
the same state in the output beam as in the input one. This means that
the amplitudes corresponding to the two arms are able to interference.
However, the gravitational perturbation is now registered not only by
the atomic matter-waves but also by the laser fields.  As shown in the
following, the `indirect' photonic contribution even dominates the
direct atomic one.

We will write the whole dephasing between the two arms as~:
\begin{equation}
\Phi = \Phi_\at  + \Phi_\ph 
\end{equation}
$\Phi_\at $ is the dephasing picked up by the atomic matter waves when
they propagate along the linear segments [AB], [BD], [AC] and [CD] of
the interferometer.  It has already been calculated in the previous
section.  $\Phi_\ph $ is the gravitational dephasing of the
electromagnetic phases involved in the stimulated Raman processes.  It
can be written as a sum over the optical elements~:
\begin{equation}
\Phi_\ph=\Phi_\ph [\A]-\Phi_\ph [\B]-\Phi_\ph [\C]+\Phi_\ph [\D] 
\label{phiphotons}
\end{equation}
Internal phase factors, corresponding to evolution at the different
frequencies of the two atomic ground states, do not appear in the
final expression, due to a proper account of energy conservation.  But
the same energy conservation law enforces that the photonic phase
$\Phi_\ph $ is present~: the change of momentum of the atoms on the
beam splitters is just equal to the change of momentum of the field
state.

\begin{figure}
$$
\epsfig{file=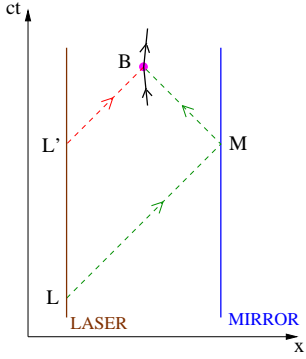,width=4cm}
$$
\caption{Space-time diagram representing the Raman processes at the
beam splitter B~: the photonic lines are the dashed lines with a slope
unity with reduced space-time variables $ct$ and $\bx $; the vertical
lines represent the motionless macroscopic objects constituting the
laser sources and the mirrors reflecting photons; the nearly vertical
heavy line represents the atoms moving with a slow velocity $v_\at \ll
c$.  The contribution $\Phi_\ph$ of the photons to the dephasing is
due to their perturbation by gravitational waves on their paths from
the laser sources to the atoms. The phases are supposed to be coherent
at the laser sources, but they are dephased differently on paths L'B
and LMB.}
\label{bs}
\end{figure}

We now evaluate the gravitational dephasing $\Phi_\ph [\B]$ of the
laser waves involved in the stimulated Raman process at beam splitter
B. They are obtained as the effect of gravitational waves on the
counter-propagating laser waves before the latter attain the atoms at
point B. We use the simplifying assumption that the Raman process is
instantaneous, so that the dephasing of the two lasers has to be
evaluated at the same spatio-temporal point. The calculation is the
same as for standard optical interferometers
\cite{Grishchuk88,Hellings92} with optical paths described on the
space-time diagram of figure (\ref{bs}).  As in the preceding section,
the macroscopic reference objects, that is the laser sources and
mirrors reflecting photons, are supposed to be at rest initially and
to stay at rest, thanks to their large mass.

The coordinates of the point B are fixed as in the preceding section.
The coordinates of the other apexes are displaced with respect to that
of B in the direction of the axis $\bx_3$ which is the propagation
direction of the lasers (see figure \ref{hyper})~:
\begin{eqnarray}
t_\M=t_\B-\tau_\MB &,&\; \bx_\M^3=\bx_\B^3+c \tau_\MB
\nonumber \\
t_\Ll=t_\B-(\tau_\MB+\tau_\LM)&,&\; \bx_\Ll^3=\bx_\B^3-c(\tau_\LM-\tau_\MB) 
\nonumber \\
t_\LL=t_\B-(\tau_\LM-\tau_\MB)&,&\; \bx_\LL^3=\bx_\Ll^3 \nonumber \\
\end{eqnarray}
We have introduced the notations $\tau_\LM$ for the time of flight of
photons from the laser source L to the mirror M and $\tau_\MB$ for the
time of flight of photons from the mirror M to the atom at B.  The
reduced velocities have simple expressions for photons~:
\begin{eqnarray}
&& \bu_\LM^3=\bu_\LLB^3=1 \qquad \bu_\MB^3=-1
\end{eqnarray}
If follows from the expressions (\ref{polariz}) that all polarization
factors involved in the evaluation of $\Phi_\ph $ have the same simple
form $\left(\beps ^{-\gamma} \mathbf{.} \bu \right)^2 = \sin^2\theta$.

The dephasing $\Phi_\ph [B]$ is then deduced from a factor
$\phi_k^{\gamma}[\B]$ representing a decomposition in momentum space~:
\begin{eqnarray}
&&\phi_k^{\gamma}[\B]=\frac{i\Omega_\ph }{2\sqrt{2}\omega} \psi_k 
e^{-i\omega\eta_\B} \nonumber \\
&&\psi_k = \bp \bm \left( e^{i\omega\tau_\MB \bp}\frac{1-e^{i\omega
\tau_\LM\bm}}{\bm}\right. \nonumber\\
&&\left. + \frac{1-e^{i\omega\tau_\MB\bp}}{\bp} 
- \frac{1-e^{i\omega(\tau_\LM-\tau_\MB)\bm}}{\bm} \right)
\nonumber \\
&&\omega\eta_\B = k_\mu x^\mu_\B = \omega t_\B - \bk \mathbf{.} \bx_\B 
\nonumber \\
&&\beta_{\pm}= 1 \pm \cos\theta \qquad \bp \bm = \sin^2\theta 
\end{eqnarray}
{}From now on, we neglect the difference between the two laser
frequencies $\Omega_1 \simeq \Omega_2=\Omega_\ph $. We introduce a
notation $\eta_\B$ to represent the phase time of the mode $k$ at the
point B.  We also introduce the notations $\beta_{\pm}$ for the
factors measuring the angle of propagation between gravitational and
electromagnetic waves. Since these waves propagate at the same speed,
potential resonances might occur in the limit of colinear propagation
$\bp=0$ or $\bm=0$, as shown by the appearance of the denominators in
$\psi_k$. In fact, these resonances do not occur because the
numerators of the fractions and, also, the polarization factor $\bp
\bm$ vanish in this limit \cite{Mashhoon80}.

We then perform the same evaluation for the other contributions,
$\phi_k^{\gamma}[\A]$, $\phi_k^{\gamma}[\C]$ and $\phi_k^{\gamma}[\D]$
which are involved in the whole photonic dephasing
(\ref{phiphotons}). Assuming for simplicity that the geometry is the
same for the four beam splitters and mirrors, these 4 terms only differ
through a global phase and their interference leads to~:
\begin{eqnarray}
\phi_k^{\gamma} &=& \frac{i\Omega_\ph }{2\sqrt{2}\omega} \psi_k \Psi_k 
\nonumber \\
\Psi_k &=& e^{-i\omega\eta_A}-e^{-i\omega\eta_B}-e^{-i\omega\eta_C}+
e^{-i\omega\eta_D} \label{phikgam}
\end{eqnarray}
The phase-times $\eta_\A$, $\eta_\C$ and $\eta_\D$ are defined as
$\eta_\B$ from the phases of the gravitational mode $k$ at the
spacetime points corresponding to the passage of the atom at
corresponding optical elements.

The magnitude of the photonic dephasing $\Phi_\ph$ is mainly
determined by the laser frequency $\Omega_\ph$ whereas the magnitude
of the atomic dephasing $\Phi_\at$ was proportional to the frequency
$\Omega_\at$. Since the latter frequency is much smaller than the
former, it is expected that~:
\begin{eqnarray}
&&\Phi_\at \ll \Phi_\ph 
\end{eqnarray}
We will see at the end of the present calculation that this is the
case. As a consequence, the correlation between atomic and photonic dephasings
will also have a negligible contribution. It follows that the phase
noise variance (\ref{eqfond}) will be determined essentially by the
photonic contribution~:
\begin{eqnarray}
\cA_\ph [\omega] &=& \frac{\Omega_\ph ^2}{4\omega^2} \frac{5}{2}
\left< \left| \psi_k \right|^2 \left| \Psi_k \right|^2 \right>_{\bn }
\label{Aphot}
\end{eqnarray}
Note that the two polarizations have the same contribution to the
present result.

We use the function (\ref{definif}) to express the squared
amplitudes~:
\begin{eqnarray}
\left| \Psi_k \right|^2 &=& 
f(\omega\eta_\AB) + f(\omega\eta_\AC) 
+ f(\omega\eta_\BD)+ f(\omega\eta_\CD)  \nonumber \\
&&\qquad- f(\omega\eta_\AD) - f(\omega\eta_\BC) 
\nonumber \\
\eta_\AB &=& \eta_\B-\eta_\A = \tau_\AB 
\left( 1 - \sin\alpha\cos\theta-\cos\alpha\sin\theta\cos\varphi\right)
\nonumber \\
\eta_\AC &=& \eta_\C-\eta_\A = \tau_\AB
\left( 1 + \sin\alpha\cos\theta-\cos\alpha\sin\theta\cos\varphi\right)
\nonumber \\
\eta_\CD &=& \eta_\AB \quad , \quad \eta_\BD = \eta_\AC 
\nonumber \\
\eta_\AD &=& \eta_\D-\eta_\A = \eta_\AB + \eta_\AC \nonumber \\
\eta_\BC &=& \eta_\C-\eta_\B = \eta_\AC - \eta_\AB 
\label{squaredPsi}
\end{eqnarray}
and~:
\begin{eqnarray}
\left| \psi_k \right|^2&=&\bp (\bp-\bm)\big[ f\big(\omega
((\tau_\LM -\tau_\MB )\bm)\big)
\nonumber \\
&& +f\big(\omega \tau_\LM \bm\big) +f\big(\omega \tau_\MB \bp\big) 
\nonumber \\
&& -f\big(\omega((\tau_\LM -\tau_\MB )\bm-\tau_\MB \bp)\big)\nonumber \\
&&\!\!\!\!\!\!\!\!\!-f\big(\omega(\tau_\LM \bm+\tau_\MB \bp)\big) \big] 
+\bp^2 f\big(\omega \tau_\MB (\bp+\bm)\big)\nonumber \\
&& +\bm (\bm-\bp) f\big(\omega \tau_\MB \bp \big)
\label{squaredpsi}
\end{eqnarray}
These equations give the phase noise variance due to photons for an
arbitrary noise spectrum $S_h [\omega]$.

In order to go further, we perform the same approximations as in the
preceding section.  We consider the case of a thermal bath with the
noise spectrum $S_h$ constant over the frequency domain relevant for
the integral (\ref{eqfond}).  We focus our attention on the limit of a
small $\Gamma$ and use the integral (\ref{propff}) as well as the
further properties of the function $f$~:
\begin{eqnarray} 
&&f(x) f(y) = 2 f(x) + 2 f(y) -f(x+y) - f(x-y)  \nonumber \\
&&\int\frac{\dd\omega}{4\pi}\frac{f(\omega \eta)f(\omega \tau)}{\omega^2}=
\Min(|\eta|,|\tau|)
\end{eqnarray}
We deduce from (\ref{squaredPsi},\ref{squaredpsi})~:
\begin{align}
&\frac{\Delta\Phi_\ph^2}{2}\simeq\frac{4}{\pi} \Omega_\ph ^2 
S_h\tau_\ph \nonumber \\
&\tau_\ph=\frac{5\pi}{32} 
\big< 2 T(\eta_\AB) + 2 T(\eta_\AC)-T(\eta_\AD)-T(\eta_\BC)\big>_\bn
\end{align}
$\tau_\ph $ has been defined so that the expression of
$\Delta\Phi_\ph^2$ has the same form as $\Delta\Phi_\at^2$ in
(\ref{decoh}); it is obtained from the auxiliary function $T$ which
has the same structure as in (\ref{squaredpsi})~:
\begin{eqnarray}
T(\eta) &=& \bp (\bp-\bm) \big[ \Min\big( |\eta|, |\tau_\LM -\tau_\MB| 
\bm) \big)\nonumber \\
&& + \Min\big( |\eta|, \tau_\LM \bm \big) 
 + \Min\big( |\eta|, \tau_\MB \bp \big)
\nonumber \\
&&  - \Min\big( |\eta|, |(\tau_\LM -\tau_\MB )\bm-\tau_\MB \bp| \big)
\nonumber \\
&&  - \Min\big( |\eta|, \tau_\LM \bm+\tau_\MB \bp \big) \big]
\nonumber \\
&& +\bp^2 \Min\big( |\eta|, \tau_\MB (\bp+\bm) \big) \nonumber \\
&& +\bm (\bm-\bp) \Min\big( |\eta|, \tau_\MB \bp) \big)
\end{eqnarray}

\begin{figure}
$$
\psfig{file=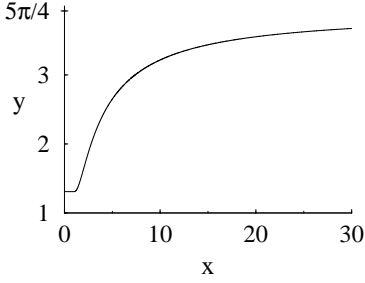,width=5cm}
$$
\caption{Variation of the ratio $y=\frac{\tau_\ph}{\tau_\MB}$ as a
function of the ratio $x=\frac{\tau_\LM}{\tau_\MB}$~; $y$ varies from
the value $\frac{5\pi}{12}$ for $x \le 1$ to the value $\frac{5\pi}{4}$
at the limit $x \gg 1$.}
\label{functionT}
\end{figure}

The lengths are of the same order for atomic and photonic lines, but
the velocity of light $c$ is much larger than the atomic velocity
$v_\at$. Hence the atomic time of flight $\tau_\AB$ is much larger
than the photonic ones $\tau_\MB$ and $\tau_\LM$.  It follows that
$\eta$ is much larger than the other time parameter appearing in the
function $\Min$ in the preceding equation, except for specific
gravitational modes propagating along the segments of the atomic
rhomb.  Disregarding these exceptions which have a negligible
contribution to the spatial mean values, we obtain a good
approximation for the function $T$~:
\begin{align}
&T(\eta)\simeq\bp (\bp-\bm)\big[ |\tau_\LM -\tau_\MB|\bm\nonumber \\
&-|(\tau_\LM -\tau_\MB )\bm - \tau_\MB \bp| \big] +\bp (\bp^2+\bm^2) 
\tau_\MB  
\end{align}
Since the function $T$ no longer depends on the parameter $\eta$, the
equivalent photonic interaction time $\tau_\ph$ is simply~:
\begin{eqnarray}
&&\tau_\ph = \frac{5\pi}{16} \big< T \big>_\bn = y \tau_\MB
\nonumber \\
&& y = \frac{5\pi}{16} \big< \bp (\bp-\bm) 
\left( |x-1| \bm - |x \bm - 2| \right) \nonumber \\
&&\qquad\qquad\qquad\qquad+ \bp (\bp^2+\bm^2) \big>_\bn 
\nonumber \\
&&x = \frac{\tau_\LM}{\tau_\MB}
\end{eqnarray}
The numerical factor $y=\frac{\tau_\ph}{\tau_\MB}$ is a function of
the ratio $x=\frac{\tau_\LM}{\tau_\MB}$ which can be obtained through
a numerical integration~:
\begin{eqnarray}
y(x) &=& \left\{ 
\begin{array}{lll}
\frac{5\pi}{2} \left( \frac{1}{2} - \frac{3x^2-3x+1}{3x^3} \right) & 
\mathrm{for} & x \geq 1 \\
\frac{5\pi}{12} & \mathrm{for} & x\leq 1
\end{array} 
\right. 
\end{eqnarray}
The function $y(x)$ is drawn on figure \ref{functionT}.

Finally the phase noise variance $\Delta\Phi_\ph ^2$ corresponding to
the photonic lines is essentially determined by the time of flight
$\tau_\MB$ of photons between atoms and mirrors~:
\begin{eqnarray}
&& \frac{\Delta\Phi_\ph^2}{2} \simeq \frac{4}{\pi} \Omega_\ph ^2 S_h y \tau_\MB  
\label{finalDeltaPhi}
\end{eqnarray}
Using the numbers corresponding to HYPER~:
\begin{eqnarray}
&&\Omega_\ph \simeq 2 \times 10^{15} \: \mbox{rad.s}^{-1} \quad , \quad 
\tau_\MB  \simeq 10^{-9} \: \mbox{s} \nonumber \\
&& \tau_\LM  \sim 3 \, \tau_\MB \quad , \quad
\tau_\ph \sim 2 \, \tau_\MB 
\end{eqnarray}
we get the following estimation of the phase noise variance associated
with photonic lines~:
\begin{equation}
\frac{\Delta\Phi_\ph^2}{2} \simeq 10^{-12}
\end{equation}
This indirect contribution due to the gravitational perturbation of
electromagnetic waves is much larger than the direct effect of
gravitational waves on atomic matter waves. This result was
anticipated from the fact that the laser frequency $\Omega_\ph$ is
much larger than the corresponding atomic quantity $\Omega_\at$.  It
was however necessary to perform the whole calculation to reach an
unambiguous conclusion, because the variance also depends on the times
of exposition of photons to the gravitational interaction, these times
$\tau_\MB$ and $\tau_\LM$ being smaller than $\tau_\AB$.

The final result proves that the decoherence of HYPER-like
interferometers is dominated by the photonic contribution.  In other
words, as far as the coupling to gravitational waves is concerned,
such interferometers essentially behave as optical detectors with a
readout mediated by atomic waves.  A second important consequence of
our result is that the photonic contribution to decoherence, though
much larger than the atomic one, is still completely
negligible. Hence, the fringe contrast of HYPER is unaffected by the
scattering of gravitational waves.

It is worth explaining this result in more details in terms of
spectra. To this aim, we come back to the expression (\ref{eqfond})
where the variance is the integral of a phase noise spectrum $S_\Phi
[\omega]$.  We then simplify this expression by performing the already
discussed approximations~: $S_h$ is considered as constant and the
crossed terms in the squared amplitudes $\left| \Psi_k \right|^2$ are
disregarded. Keeping only the square terms we get a typical
dependence~:
\begin{eqnarray}
S_\Phi [\omega] &\sim& S_h \Omega_\ph ^2
\frac{f(\omega\tau)}{\omega^2} 
\end{eqnarray}
where $\tau$ is one of the times of flight involved in equation
(\ref{squaredpsi}). The phase noise spectrum has a magnitude of the
order of $S_h \Omega_\ph^2 \tau^2$ and a bandwidth of the order of
$\tau^{-1}$ which lead to an integral of the order of $S_h
\Omega_\ph^2 \tau$. This simple estimate fits the result
(\ref{finalDeltaPhi}) obtained through a more rigorous calculation.

Now the phase noise level may be measured as an equivalent vibration
noise for the mirrors reflecting the lasers. This equivalent noise,
written in terms of the position $q$ of a mirror, is approximated as~:
\begin{eqnarray}
&&S_q [\omega] \sim S_h (c \tau)^2 
\sim 10^{-34} \left( \mbox{m} / \sqrt{\mbox{Hz}}\right)^2 
\end{eqnarray}
We have again used the numbers of HYPER with $c\tau$ nearly equal to
1m.  This corresponds to a noise level $\sqrt{S_q} \sim 10^{-17}
\mbox{m} / \sqrt{\mbox{Hz}}$ which is far beyond the vibration noise
level $\sqrt{S_q} \sim 10^{-12} \mbox{m} / \sqrt{\mbox{Hz}}$ which is
the target of the HYPER instrument. This discussion confirms that the
phase noise induced by the scattering of gravitational waves is
completely negligible in HYPER-like interferometers. In particular, it
shows that this fundamental spacetime noise is smaller than the
residual phase noise corresponding to mechanical vibrations of the
mirrors.

\section{Discussion}

It had been suggested that atomic interferometers could be sensitive
to a decoherence effect stemming from intrinsic spacetime
fluctuations. In the present paper, we have studied the effect
associated with the scattering of gravitational waves, which we expect
to be the dominant source of spacetime fluctuations in the frequency
domain of interest for atomic probes. Taking the numbers of the
project HYPER as an example, we have shown that this effect is
completely negligible. Essentially, this result has to be considered
as positive for the project HYPER~: as a matter of fact, it entails
that phase shifts stemming from the Lense Thirring effect, the
observation of which constitutes the main scientific objective of the
project \cite{Hyper00}, will not be washed out by the stochastic
background of gravitational waves.

The results could be different when considering larger sources of
spacetime fluctuations or different couplings to matter
\cite{Percival97,PercivalS97,Power00,Amelino99,Amelino00}.  But it is
natural to think that the gravitational waves, which are predicted to
exist in our environment by general relativity, are the dominant
source of spacetime fluctuations in the frequency range of
interest. In any case, the results derived in this paper from well
established knowledge about gravitational waves and their interaction
with matter may be used as a reference point to which more speculative
proposals have to be compared.

In this concluding section, we discuss a few points which may be
relevant for a larger class of atomic interferometry experiments.  In
order to discuss the scaling properties of the decoherence effect with
respect to the main relevant parameters, we rewrite the atomic and
photonic contributions~:
\begin{eqnarray}
&&\Delta\Phi_\at ^2 \sim S_h \, \Omega_\at^2  \sin^2 (2\alpha)
\, \tau_\at  \nonumber \\
&&\Delta\Phi_\ph ^2 \sim S_h \, \Omega_\ph  ^2 \, \tau_\ph 
\label{finalvariances}
\end{eqnarray}
$S_h$ is the gravitational noise spectrum, supposed to be constant
over the frequency range of interest, $\Omega_\at$ and $\Omega_\ph$
are the kinetic energy of the probes measured as a frequency,
$\tau_\at$ and $\tau_\ph$ represent the times of flight of atoms and
photons respectively.

The phase noise variances scale as the times of exposition of the
probe to the gravitational perturbation, which means that the effect
can be understood as resulting from a Brownian-like diffusion due to
stochastic fluctuations of spacetime \cite{Percival97}. It is worth
noticing that this result is directly linked to the assumption of a
flat noise spectrum. A different scaling law for the spectrum, such as
that predicted by most cosmological models, would necessarily lead to
a different dependence of $\Delta\Phi^2$ versus the time of
exposition.

Now an important point has to be emphasized at this stage~: contrarily
to what could have been expected, the key atomic parameter which
determines decoherence is not the rest energy $mc^2$ of the atomic
probe but rather the kinetic energy $mv_\at^2/2$. For atoms, this
makes a significant difference in the evaluation of $\Delta\Phi^2$
which scales as the square of $\Omega_\at$. For photons, the mass
vanishes but the kinetic energy $\Omega_\ph$ is merely the
frequency. This result is directly related to qualitative arguments
already presented~: the phase noise can be expressed through gauge
invariant expressions and, as a consequence, it may be evaluated in a
specific gauge, for example the TT gauge; in this gauge, gravitational
waves have no effect on massive particules at rest. Hence,
matter-waves corresponding to slow atoms are poorly coupled to
gravitational waves. Clearly, this is not true for the detection of
quasi-static gravitational field such as the Lense-Thirring effect
looked for in the HYPER project \cite{Hyper00}. For such a
measurement, the sensitivity is effectively determined essentially by
the rest mass frequency, which explains why atomic interferometers may
be used as highly sensitive probes of quasistatic metric effects
\cite{Peters01,Borde00,Salomon01}.

Incidentally, these discussions imply that cold atoms are poorly
adapted to the detection of gravitational waves. Should we aim at
observing an effect of stochastic gravitational backgrounds on
interferometers, more natural strategies would use either optical
interferometers or atoms with as high a kinetic energy as possible.
If the beam splitters and mirrors are built up with stimulated Raman
process, this raises the problem that the transverse velocity and,
therefore, the area of the rhomb decrease when the kinetic energy is
increased.

Finally, we want to come back to the qualitative arguments bearing on
Planck units which have been evoked in the beginning of this paper.
To this aim, we introduce new characterizations of the gravitational
noise spectrum~:
\begin{equation}
S_h [\omega]= \frac{16G}{5c^5} \, \kB T_{\gw} [\omega] =
\frac{16G}{5c^5} \, \hbar \omega n_\gw[\omega]
\end{equation}
$n_\gw$ is the number of gravitons per mode and $T_{\gw}$ is an
effective noise temperature associated with the gravitational
background. Both quantitites depend on the frequency in general.  For
the binary confusion background considered above, we obtain the
following values~:
\begin{eqnarray}
10^{-6} \: \mbox{Hz} <&\frac{\omega}{2\pi}&< 10^{-4} \: \mbox{Hz}
\nonumber \\ 
S_h \sim 10^{-34} \: \mbox{Hz}^{-1} \quad&,&\quad
T_{\gw} \sim 10^{41} \: \mbox{K} 
\end{eqnarray}
The effective noise temperature has an extremely high value, even
higher than Planck temperature $\sim 10^{32}$K. This emphasizes the
unconventional character of the noise temperature $T_\gw$ from the
point of view of thermodynamics. In fact, gravitational waves interact
so weakly with matter that the associated thermalization time is
extremely long.

We also introduce a parameter $\Theta_\gw$ which measures the noise
temperature as a frequency~:
\begin{eqnarray}
\Theta_\gw &=& \pi \frac{\kB T_{\gw}}{\hbar} 
= \pi \omega n_\gw[\omega]\sim 3 \times10^{52} \: \mbox{s}^{-1} 
\end{eqnarray}
Using this parameter, we rewrite the phase noise variances
(\ref{finalvariances}) as~:
\begin{eqnarray}
&&\Delta\Phi_\at ^2 \sim \left( \Omega_\at \tP \right)^2 \sin^2(2\alpha) \,
\Theta_\gw \tau_\at  
\nonumber \\
&&\Delta\Phi_\ph ^2 \sim \left( \Omega_\ph  \tP \right)^2 \Theta_\gw \tau_\ph 
\nonumber \\
&& \tP = \sqrt{\frac{G\hbar }{c^5}} \sim 10^{-43} \: \mbox{s}
\end{eqnarray}
We still notice the linear dependence of the variances with respect to
the time of interaction, which is characteristic of a Brownian-like
diffusion process, as already discussed. We also observe the quadratic
dependence of the same quantities in the Planck time $\tP$, which just
means that we are dealing with effects linear in the Newton constant
$G$.  Besides these two time parameters, the variances depend on two
frequency parameters $\Omega$ and $\Theta_{\gw}$ which measures
respectively the kinetic energy of the probe and the effective noise
temperature of the gravitational background.

In the case of the atomic probe, the phase noise variance may
equivalently be written in terms of the Planck mass~:
\begin{eqnarray}
&&\Delta\Phi_\at ^2 \sim 
\left( \frac{m v_\at ^2}{\mP c^2} \right)^2 \sin^2(2\alpha) \, 
\Theta_{\gw}  \tau_\at  
\end{eqnarray}
The fraction $\frac{mv_\at^2}{\mP c^2}$ illustrates the simple scaling
argument presented in the introduction~: for microscopic masses, this
fraction is much smaller than unity so that the coupling to
gravitational fluctuations tends to become negligible. This result has
to be contrasted to the fact that the scattering of gravitational
waves tends to become the dominant source of decoherence for
macroscopic motions \cite{Reynaud01,Reynaud02}. It is however worth
acknowledging that the fraction $\frac{mv_\at^2}{\mP c^2}$ is only one
of the factors which determine the phase noise variance
$\Delta\Phi_\at ^2$. It is therefore necessary to perform the whole
calculation, as we did in the present paper for atomic
interferometers, before reaching a quantitative conclusion about the
effect of gravitational fluctuations on decoherence.


\begin{thebibliography}{99}
\bibitem{Feynman99}  Feynman R.P., Moringo F.B., Wagner W.G. and Hatfield B.,
\textit{Feynman Lectures on Gravitation} (Penguin, 1999). 

\bibitem{Feynman63}  Feynman R.P., 
{Acta Physica Polonica} \textbf{24} (1963) 711.

\bibitem{Karolyhazy66}  Karolyhazy F., {Nuovo Cim.} \textbf{42A} 
(1966) 390.

\bibitem{Diosi89}  Diosi L., {Phys. Rev.} \textbf{A40} (1989) 1165.

\bibitem{Penrose96}  Penrose R., {Gen. Rel. Grav.} \textbf{28} 
(1996) 581.

\bibitem{Ellis98}  Ellis J., Kanti P., Mavromatos N.E., Winstanley E., 
Nanopoulos D.V., {Mod. Phys. Lett.} \textbf{A13} (1990) 303.

\bibitem{Percival97}  
Percival I.C., {Phys. World} \textbf{10} March issue (1997)
48.

\bibitem{PercivalS97}  Percival I.C. and Strunz W.T., 
{Proc. R. Soc. London} \textbf{A453} (1997) 431.

\bibitem{Power00}  Power W.L. and Percival I.C., 
{Proc. R. Soc. London} \textbf{A455} (2000) 991.

\bibitem{Amelino99}  Amelino-Camelia G., {Phys. Rev.} \textbf{D62}
(1999) 024015.

\bibitem{Amelino00}  Amelino-Camelia G., 
{Nature} \textbf{408} (2000) 661.

\bibitem{Peters99}  Peters A., Chung K.Y. and Chu S.,
{Nature} \textbf{400} (1999) 849.

\bibitem{Peters01}  Peters A., Chung K.Y. and Chu S.,
{Metrologia} \textbf{38} (2001) 25.

\bibitem{Gustavson00}  Gustavson T.L., Landragin A. and Kasevich M.A.,
{Class. Quant. Gravity} \textbf{17} (2000) 2385.

\bibitem{Hyper00}  \textit{HYPER, Hyper-precision cold atom interferometry
in space}, Assessment study report \textbf{10} (ESA-SCI, 2000).

\bibitem{Misner}  Misner C.W., Thorne K.S. and Wheeler J.A., 
\textit{Gravitation} (Freeman and Company, 1973).

\bibitem{Will90}  Will C.M., {Science} \textbf{250} (1990) 770.

\bibitem{Damour94}  Damour T. in \textit{Gravitation and Quantizations},
B. Julia and J. Zinn-Justin eds, (North Holland, 1994).

\bibitem{Weinberg65}  Weinberg S., {Phys. Rev.} \textbf{138} 
(1965) B988.

\bibitem{Grishchuk77}  Grishchuk L.P., {Usp. Fiz. Nauk} \textbf{121}
(1977) 629.

\bibitem{Zeldovich86}  Zeldovich Ya.B. and Grishchuk L.P., {Usp. Fiz.
Nauk} \textbf{149} (1986) 695.

\bibitem{Schutz99}  Schutz B., {Class. Quant. Grav.} \textbf{16} 
(1999) A131.

\bibitem{Maggiore00}  Maggiore M., {Physics Reports} \textbf{331} 
(2000) 283.

\bibitem{Ungarelli00}  Ungarelli C. and Vecchio A., 
{Phys. Rev.} \textbf{D63} (2001) 064030.

\bibitem{Hils90}  Hils D., Bender P.L. and Webbink R.F., {Astrophys.
J.} \textbf{360} (1990) 75.

\bibitem{Giazotto97}  Giazotto A., Bonazzola S. and Gourgoulhon E., 
{Phys. Rev.} \textbf{D55} (1997) 2014.

\bibitem{AtomInt97}  \textit{Atom interferometry}, P. Berman ed., 
(Academic Press, 1997). 

\bibitem{Zeh70}  
Zeh H.D., {Found. Phys.} \textbf{1} (1970) 69.

\bibitem{Dekker77}  
Dekker H., {Phys. Rev.} \textbf{A16} (1977) 2126.

\bibitem{Zurek81}  
Zurek A.J., {Phys. Rev.} \textbf{D24} (1981) 1516.

\bibitem{Zurek82}  
Zurek A.J., {Phys. Rev.} \textbf{D26} (1982) 1862.

\bibitem{Caldeira83}  Caldeira A.O. and Leggett A.J.,
{Physica} \textbf{121 A} (1983) 587. 

\bibitem{Caldeira85}  Caldeira A.O. and Leggett A.J.,
{Phys. Rev.} \textbf{A31} (1985) 1059.

\bibitem{Joos85}  
Joos E. and Zeh H.D., {Z. Phys.} \textbf{B59} (1985) 223.

\bibitem{Raimond01}  Raimond J.M., Brune M. and Haroche S., 
{Rev. Mod. Phys.} \textbf{73} (2001) 565.

\bibitem{Imry90}  Stern A., Aharonov Y., Imry Y., {Phys. Rev.} 
\textbf{A41} (1990) 3436.

\bibitem{Reynaud01}  Reynaud S., Maia Neto P.A., Lambrecht A. and Jaekel
M.T., {Europhys. Lett.} \textbf{54} (2001) 135.

\bibitem{Reynaud02}  Reynaud S., Maia Neto P.A., Lambrecht A., Jaekel M.T. 
and Lamine B., to appear in {Int. J. Mod. Phys.}, 
\textbf{A} (2002) [arXiv:gr-qc/0111105]. 

\bibitem{Blanchet00}  Blanchet L., Kopeikin S. and Sch\"afer G., 
in \textit{Gyros, Clocks, and Interferometers: Testing Relativistic Gravity in 
Space}, C. L\"ammerzahl, C.W.F. Everitt, F.W. Hehl eds., Lecture Notes in
Physics \textbf{562} (2001) 141 [arXiv:gr-qc/0008074]. 

\bibitem{Grishchuk90}  Grishchuk L.P. and Sidorov Y.V., {Phys. Rev.}
\textbf{D42} (1990) 3413.

\bibitem{Jaekel94}  Jaekel M.T. and Reynaud S., {Phys. Lett.} 
\textbf{A185} (1994) 143.

\bibitem{Jaekel95QSO}  Jaekel M.T. and Reynaud S., 
{Quant. Semicl. Optics} \textbf{7} (1995) 639.

\bibitem{Jaekel95AP}  Jaekel M.T. and Reynaud S., {Ann. Physik} 
\textbf{4} (1995) 68.

\bibitem{Landau} Landau L.D. and Lifshitz E.M., \textit{Course of
theoretical physics: The classical theory of fields} (Butterworth
Heinemann, 4th revised edition 1998) \S 94.

\bibitem{Grishchuk88}  Grishchuk L.P., {Usp. Fiz. Nauk} 
\textbf{156} (1988) 297.

\bibitem{Hellings92}  Hellings R.W., in \textit{Detection of gravitational 
waves}, D. Blair ed. (Cambridge University Press, 1992), 453.

\bibitem{Mashhoon80}  Mashhoon B. and Grishchuk L.P., {Astrophys. J.} 
\textbf{236} (1980) 990.

\bibitem{Borde89}  Bord\'e C.J., {Phys. Lett.} \textbf{140} (1989)
10.

\bibitem{Borde92}  Bord\'e C.J., in \textit{Laser Spectroscopy X},
M. Ducloy, E. Giacobino and G. Camy eds., (World Scientific, 1992) 239.

\bibitem{Kasevich92}  Kasevich M. and Chu S. {App. Phys.} \textbf{B54} 
(1992) 321.

\bibitem{Weiss94}  Weiss D.S., Young B.C. and Chu S. {App. Phys.} 
\textbf{B59} (1994) 217.

\bibitem{Storey94}  Storey P. and Cohen-Tannoudji C., {J. Physique} 
\textbf{II-4} 
(1994) 1999.

\bibitem{Borde00}  Bord\'e C.J., Houard J.-C. and Karasiewicz A., in 
\textit{Gyros, Clocks, and Interferometers: Testing Relativistic Gravity in 
Space}, C. L\"ammerzahl, C.W.F. Everitt, F.W. Hehl eds., Lecture Notes in
Physics \textbf{562} (2001) 405.

\bibitem{Salomon01}  Salomon C., Dimarcq N., Abgrall M. \textit{et al}, 
{Comptes-Rendus de l'Acad\'emie des Sciences}, \textbf{2-IV} (2001) 1313.

\end{thebibliography}
\end{document}